\documentclass[fleqn,usenatbib]{hclass}
\usepackage{newtxtext,newtxmath}

\usepackage[T1]{fontenc}
\usepackage{ae,aecompl}
\usepackage{graphicx}	
\usepackage{amsmath}	
\usepackage{amssymb}	
\usepackage{slashbox}
\usepackage{pdflscape}
\newcommand{\sty}{\scriptstyle}
\newcommand{\ssty}{\scriptscriptstyle}
\newcommand{\tsty}{\textstyle}

\newcommand{\be}{\begin{equation}}
\newcommand{\ee}{\end{equation}}
\newcommand{\lb}[1]{\label{#1}}
\newcommand{\dl}{d_{\ssty L}}

\newcommand{\dg}{d_{\ssty G}}
\newcommand{\dz}{d_z}
\newcommand{\gaml}{\gamma_{\ssty L}}

\newcommand{\gamg}{\gamma_{\ssty G}}
\newcommand{\gamz}{\gamma_{\ssty z}}

\newcommand{\dobs}{d_{{\tsty {\ssty \rm obs}}}}
\newcommand{\Nobs}{N_{{\tsty {\ssty \rm obs}}}}
\newcommand{\Vobs}{V_{{\tsty {\ssty \rm obs}}}}
\newcommand{\gobs}{\gamma_{{\tsty {\ssty \rm obs}}}}

\title[Galaxy Distributions as Fractal Systems]
      {Galaxy Distributions as Fractal Systems}
\author[S.\ Teles, A.\ R.\ Lopes \& M.\ B.\ Ribeiro]
{
Sharon Teles,$^{1}$$^{\mbox{\thanks{E-mail: steles.ts@gmail.com;
Orcid: 0000-0003-4497-9161}}}$
Amanda R. Lopes,$^{2,3}$$^{\mbox{\thanks{E-mail: amandalopes1920@gmail.com;
Orcid: 0000-0002-6164-5051}}}$
and Marcelo B. Ribeiro$^{4}$$^{\mbox{\thanks{E-mail: mbr@if.ufrj.br;
Orcid: 0000-0002-6919-2624}\thanks{Corresponding author}}}$
\\
$^{1}$Valongo Observatory, Universidade Federal do Rio de Janeiro, Rio de
      Janeiro, Brazil\\
$^{2}$Department of Astronomy, Observat\'{o}rio Nacional, Rio de Janeiro,
      Brazil\\
$^{3}$Instituto de Astrof\'{\i}sica de La Plata, CONICET--UNLP, La Plata,
      Argentina\\
$^{4}$Physics Institute, Universidade Federal do Rio de Janeiro, Rio de
      Janeiro, Brazil
}
\pubyear{2022}
\begin{document}
\label{firstpage}
\pagerange{\pageref{firstpage}--\pageref{lastpage}}
\maketitle
\begin{abstract}
This paper discusses if large scale galaxy distribution samples containing
almost one million objects can be characterized as fractal systems. The 
analysis performed by \citet{teles2020} on the UltraVISTA DR1 survey is
extended here to the SPLASH and COSMOS2015 catalogs, hence adding
750k new galaxies with measured redshifts to the studied samples. The
standard $\Lambda$CDM cosmology having $H_0=(70\pm5)$ km/s/Mpc and number
density tools required for describing these galaxy distributions as single
fractal systems with dimension $D$ are adopted. We use the luminosity
distance $\dl$, redshift distance $\dz$ and galaxy area distance
(transverse comoving distance) $\dg$ as relativistic distance definitions
to derive galaxy number densities in the redshift interval $0.1\le z\le4$
at volume limited subsamples defined by absolute magnitudes in the
K-band.  Similar to the findings of \citet{teles2020}, the results show
two consecutive redshift scales where galaxy distribution data behave as
single fractal structures. For $z<1$ we found $D=1.00\pm0.12$ for the
SPLASH galaxies, and $D=1,39\pm0.19$ for the COSMOS2015. For $1\le z\le4$
we respectively found $D=0.83^{+0.36}_{-0.37}$ and $D=0.54^{+0.27}_{-0.26}$.
These results were verified to be robust under the assumed Hubble constant
uncertainty. Calculations considering blue and red galaxies subsamples in
both surveys showed that the fractal dimensions of blue galaxies as
basically unchanged, but the ones for the red galaxies changed mostly to
smaller values, meaning that $D$ may be seen as a more intrinsic property
of the distribution of objects in the Universe, therefore allowing for the
fractal dimension to be used as a tool to study different populations of
galaxies. All results confirm the decades old theoretical prediction of a
decrease in the fractal dimension for $z>1$.
\end{abstract}
\section{Introduction}\lb{intro}

\textit{Fractal analysis} of the galaxy distribution consists of
applying the standard techniques of fractal geometry to a given
galaxy redshift survey dataset with the aim of determining if this
distribution has fractal features. In other words, the goal is to
test the \textit{fractal galaxy distribution hypothesis}, that is,
the assumption that this distribution can be described as a fractal
system. This analysis is done by calculating the key feature of
fractal systems, the \textit{fractal dimension} $D$, which basically
characterizes the distribution's irregularity \citep{mandelbrot83}.
In the context of large-scale galactic clustering $D$ basically
determines galactic clustering sparsity or, complementarily, the
dominance of voids in the distribution. If $D$ is smaller than 3,
which is the topological dimension where the fractal structure is
embedded, it means that the structure has irregular patterns.
Decreasing values of $D$ means increasing sparsity in the galactic
clustering \citep{pietronero87,coleman92,ribeiro98}. 

The simplest way to characterize a fractal system is by means
of the \textit{single fractal dimension}, since it reduces the
quantification of the irregular patterns within the system by
means of a unique value for $D$. More complex structures can also
be described by the single fractal approach, because a fractal
system may possess different single values for $D$ at different
distance ranges, that is, single fractal systems in sequence at
different data ranges \citep{sylos98}. The alternative way is
called \textit{multifractal}, where the fractal system has several
fractal dimensions in the same scaling range, that is, a spectrum
of dimensions whose maximum value corresponds to the single fractal
dimension the structure would have if it were treated as a single
fractal \citep{gabrielli2005}.

Galaxy redshift surveys datasets allow the determination of $D$ by
means of plots of observed number density vs.\ distance drawn from
volume-limited samples. However, galaxies located at redshift depths
where $z\gtrsim 0.1-0.2$ cannot provide consistent volume densities
without considering relativistic effects. That happens because
relativistic cosmological models possess several distance definitions
\citep{ellis71,ellis2007,holanda2010} and at those redshift ranges a
single empirically determined value for $z$ corresponds to different
distance values. Moreover, in relativistic cosmology the geometrical
locus of astronomical observations lies along the past light cone,
which means that even spatially homogeneous cosmological models like
the standard Friedmann-Lema\^{\i}tre-Robertson-Walker (FLRW) are
characterized by observational inhomogeneities at high redshift ranges
\citep{juracy2008}, since any distance measure required in the
determination of volume densities will necessarily depart the local
spatially homogeneous hypersurfaces of these models at $z\gtrsim
0.1-0.2$ \citep{ribeiro92b,ribeiro95}. Hence, relativistic effects
need to be taken into account when fractal analyses are performed
from moderate to high redshift ranges \citep{ribeiro2001b}.

The discussion above summarizes the scope of \textit{fractal cosmology},
as consisting of modeling the large-scale structure of the Universe by
assuming that its galaxy distribution behaves as a fractal pattern. It
has been previously known in the literature as \textit{hierarchical
cosmology} due to discussions regarding the possible hierarchical
structuring of the Universe in the beginning of the 20th century
\citetext{\citealp{charlier08,charlier22}; \citealp{einstein22};
\citealp{selety22}; \citealp{amoroso29}; \citealp{astrobio14}, pp.\
410--412, 1975--1976}. Attempts to theoretically describe and
empirically characterize this hierarchical galaxy structure were also
proposed \citetext{\citealp{carpenter38}; \citealp{vaucouleurs60,
vaucouleurs70}; \citealp{wertz70,wertz71}; \citealp{haggerty72};
\citealp{astrobio14}, pp.\ 371--372, 572--574}, nevertheless, after the
appearance of fractal geometry in the 1980s it became clear that the
old hierarchical cosmology concepts are essentially the same as those
of fractal cosmology, leading in fact to the same expressions, albeit
with different terminology \citep{ribeiro94,ribeiro98}.

Early hierarchical cosmology models were proposed within the framework
of Newtonian cosmology \citep{wertz70,wertz71,haggerty72}, as well as
in more recent fractal cosmology ones \citep{elcio99,elcio2004}. Later
on relativistic fractal cosmologies considering a fractal system
embedded in a 4-dimensional spacetime along the observer's past light
cone were proposed \citep{ribeiro92a,ribeiro93,emr2001,ribeiro2001a,
ribeiro2001b,ribeiro2005}. Other authors have also discussed relativistic
cosmological models with theoretical fractal features \citep{wesson78,
mureika2004,mureika2007,sylos2011,felipe2013,hossie2018,sadri2018,cosmai2019,
jawad2019, cacciatori2021} or by assuming single fractal or multifractal
patterns in observational scenarios using Newtonian or relativistic models
\citep{jones88,martinez90,pan2000,gaite2005,gaite2007,gaite2018,gaite2019,
stahl2016,raj2019,bruno2020}.

The question of whether or not there would be a transition to homogeneity
in the galaxy distribution at some yet to be determined scale is still
observationally controversial. Some recent studies argue that the latest
galaxy distribution data indicate a transition to homogeneity
\citep{scrimgeour2012,garcia2018,gaite2021,goncalves2021}, whereas others
disagree with such a conclusion \citep{gabriela,chacon16,marzo2021,
teles2020}. This observational tension seems to be a result of sometimes
using shallow galaxy redshift data, and how data is theoretically
interpreted and statistically handled, not infrequently due to the
unwarranted assumption that relativistic effects can be ignored in fractal
cosmology studies \citep{ribeiro2001b,ribeiro2005,juracy2008}. We shall
return to this point below.

In addition, it must be mentioned the now decades old theoretical
prediction that a possible galaxy fractal structure must lead to a
decrease in the fractal dimension for $z>1$ even in spatially homogeneous
FLRW cosmologies. This is due to the fact that in these cosmologies the
volume density significantly decreases at such scales when calculated
along the past light cone, as it must be, and that inevitably leads to
a decrease in the fractal dimension $D$ beyond that range \citetext{see:
\citealp{ribeiro92b}, Fig.\ 1; \citealp{ribeiro95}, Figs.\ 1 and 3;
\citealp{ribeiro2001b}, Fig.\ 2}.

Another source for this observational tension lies on the difficulties
for testing the fractal galaxy distribution hypothesis at large-scales
due to, until recently, lack of data at $z>2$, or insufficient galaxy
numbers with measured redshifts at $1<z<2$. Nevertheless, \citet{gabriela}
were able to test this hypothesis using the FORS Deep Field (FDF) dataset
consisting of 5558 galaxies in the range $0.45\le z\le5.0$, and concluded
that at $z\lesssim1.3-1.9$ the sample presented an average single fractal
dimension of $D=1.4^{\ssty+0.7}_{\ssty-0.6}$, whereas beyond this
threshold they obtained $D=0.5^{\ssty +1.2}_{\ssty-0.4}$. This study
provided the first observational support for the above mentioned
theoretical prediction of a decreasing fractal dimension at larger scales,
even despite the relatively high data uncertainties in the measure of $D$
ensued by the indirect luminosity function method employed by the authors
to obtain volume-limited samples.

This line of investigation was further advanced by \cite{teles2020}, who
carried out a fractal analysis of the UltraVISTA DR1 survey containing
219300 measured redshift galaxies, a sample considerably larger than the
FDF one, and obtained volume-limited samples directly from measured
redshift data instead of the indirect luminosity function methodology.
This study was performed considering a FLRW cosmological model, and
led to improved results in terms of better defined threshold for
moderate and high scaling ranges, smaller uncertainties and results
more in line with each other considering all cosmological distance
definitions. They reached at conclusions similar to \cite{gabriela},
i.e., that a volume-limited subsample of the UltraVISTA DR1 galaxy
distribution can also be characterized as a fractal system with two
consecutive scaling ranges with the following median dimensions and
uncertainties: $D=\left(1.58\pm0.20\right)$ for $z<1$, and
$D=\left(0.59\pm0.28\right)$ for $1\leq z\leq 4$. These results provided
further empirical support to the early theoretical prediction of a
decrease in the fractal dimension at larger scales. 

This work aims at extending the study carried out by \citet{teles2020}
in addition to testing if the fractal dimension
changes for different galaxy type subsamples. It applies the same
methodology and underlying cosmology, but uses instead data from the
COSMOS2015 and SPLASH redshift surveys. The former galaxy catalog
considerably enlarged the UltraVISTA DR1 number of galaxies observed
in the same northern hemisphere observational field, almost tripling
the total number of objects, from 219300 to 578379, whereas the latter
has 390362 objects with measured redshifts surveyed in a portion of
the southern hemisphere. Together they added almost 750k new galaxies
up the $z=6$ in comparison to the number of objects studied by
\cite{teles2020}, providing then a considerably larger galaxy
distribution sample to perform fractal analysis.

The conclusions reached here provide further empirical support that the
galaxy distribution can be characterized by two subsequent fractal
scaling ranges at decreasing single fractal dimension values and with
no detectable transition to homogeneity up to the redshift limits of
both surveys. Two volume-limited subsamples were generated in both
surveys by means of filtering through absolute
magnitudes obtained in the K-band. For $z<1$ we obtained $D=1.00\pm0.12$
for the SPLASH galaxies, and $D=1,39 \pm0.19$ for the COSMOS2015. For
$1\leq z\leq4$ we respectively found $D=0.83^{+0.36}_{-0.37}$ and
$D=0.54^{+0.27}_{-0.26}$. These results turned out to be robust under
the adopted Hubble constant uncertainty of $H_0=(70\pm5)$ km/s/Mpc.

Further subsamples were generated by selecting blue,
star forming, galaxies and red, quiescent, ones and subsequently
filtering them through the same absolute magnitude criterion
described above. The fractal dimensions of blue galaxies turned out
either unchanged or only marginally changed as compared to the
unselected and filtered samples. However, the red galaxies had their
fractal dimensions becoming noticeably smaller in most cases, apart
from the red COSMOS1015 whose $D$ values increased, also noticeably.

Such results suggest that single fractal dimensions
may be used not only as a descriptors of galaxy distributions, but
also as tools to trace galaxy types and/or their evolutionary stages
at different redshift ranges.  Besides, all results obtained here
provide further empirical confirmation of the theoretical prediction
of a decrease in the fractal dimensions at ranges where $z>1$.

The plan of the paper is as follows. Section \ref{fractal-cos}
summarizes the essential tools of relativistic fractal geometry
required for testing the fractal galaxy distribution hypothesis. 
Section \ref{fractal-analysis} describes the observational details
of the COSMOS2015 and SPLASH redshift surveys relevant to this
work, as well as the data handling required for the application of
fractal tools to these datasets. Section \ref{results} presents the
results of the fractal analysis of the COSMOS2015 and SPLASH
galaxy distributions, comparing them with previous results reached
with the UltraVISTA DR1 and FDF surveys. Section
\ref{blue-red} presents fractal dimensions by generating blue, star
forming, and red, quiescent, galaxy subsamples of the COSMOS2015 and
SPLASH datasets, and compares the results with the ones obtained with
their respective unselected samples showed in the previous section.
Section~\ref{conclusion} presents our conclusions.

\section{Relativistic Fractal Cosmology}\lb{fractal-cos}

Fractal systems are characterized by power-laws, property known
since Mandelbrot's (\citeyear{mandelbrot83}) original studies on
fractals. Indeed, the connection of early galactic structure
observations to hierarchical cosmology, and then to fractals, was
done through the \textit{observed} power-law features of galaxy
distribution \citep{vaucouleurs70,pietronero87}. Naturally, the
relativistic fractal cosmology definitions and concepts can only
make sense if conceived in this same context, that is, power-law
relationships among \textit{observable quantities}. This section
presents a brief review of concepts and definitions appropriate
for the description of possible fractal patterns in the galaxy
distributions at moderate and large redshift scales \citep[see]
[Sec.\ 2, for more details]{teles2020}.

Let $\Vobs$ be the \textit{observational volume} defined as follows,
\be
\Vobs=\frac{4}{3} \pi {(\dobs)}^3,
\lb{vobs}
\ee
where $\dobs$ is an \textit{observational distance}. The
\textit{observed number density} $\gobs^\ast$  is given by,
\be
\gobs^\ast=\frac{\Nobs}{\Vobs},
\lb{gobs-ast}
\ee
where $\Nobs$ is the \textit{observed cumulative number counts} of
cosmological sources, that is, galaxies. It is clear that
$\gobs^\ast$ gives the number of sources per unit of observational
volume out to a distance $\dobs$, in addition to being a radial
quantity and, thus, cannot be understood in statistical sense because
it does not average all points against all points.

The \textit{Pietronero-Wertz hierarchical (fractal) cosmology model}
\citetext{\citealp{ribeiro94}, Sec.\ 3; \citealp{ribeiro98}, Sec.\
III.4} has as key underlying hypothesis a phenomenological expression
called the \textit{number-distance relation}, written as follows,
\be
\Nobs=B \, {(\dobs)}^D, 
\lb{Nobs}
\ee
where $B$ is a positive constant and $D$ is the single fractal
dimension. If in the expression above $D=3$, $\Nobs$ grows with
$\Vobs$ and galaxies would evenly distribute along all regions of
the observed space. However, if $D<3$, as $\dobs$ increases $\Nobs$
grows at a smaller pace than $\Vobs$, creating then gaps in the galactic
distribution, that is, regions devoid of galaxies. Alongside these
galactic gaps there would be regions where galaxies clump. Therefore,
voids and galactic clumpiness would be a by-product of a fractal galaxy
structure whose fractal dimension is smaller than the topological
dimension where the galaxy structure is embedded. In this scenario
the fractal dimension becomes a descriptor of galactic clumpiness
or, complementarily, the dominance of voids in the galactic structure.

One must note that $\Nobs$ is a cumulative quantity. So, if beyond a
certain distance there are no longer galaxies then $\Nobs$ no longer
increases with $\dobs$. If, on the other hand, objects are still
detected and counted, even at irregular pace, then it continues to
increase. This rate of growth can be affected by observational biases,
possibly leading to an intermittent behavior, however $\Nobs$ must
grow or remain constant and, therefore, the exponent in Eq.\
(\ref{Nobs}) must be positive or zero.

Substituting Eqs.\ (\ref{vobs}) and (\ref{Nobs}) into Eq.\
(\ref{gobs-ast}) we obtain the \textit{De Vaucouleurs density
power-law} \citep{pietronero87,ribeiro94}, 
\begin{equation}
\gobs^\ast = \frac{3B}{4\pi}{(\dobs)}^{D-3}.
\lb{gstar3}
\end{equation}
Hence, if the observed galaxy distribution is found to have $D<3$,
the observational number density above decays as a power-law. If
$D=3$ galaxies are evenly distributed and the galactic structure is
said to be \textit{observationally homogeneous}. In this case the
number density becomes constant and distance independent. Smaller
values of $D$ imply steeper power-law decays and, consequently,
more gaps in the galaxy distribution. Note that the power-law above
allows for the empirical determination of different fractal
dimensions in two or more scaling ranges dependent on the intervals
of $\dobs$.

The expressions above are directly applicable in Newtonian cosmology,
however, to use them in a relativistic setting some relativistic
concepts must come to the forefront. First, in relativistic cosmology
the geometrical locus of observations is the observer's past light
cone, which  means that that even spatially homogeneous cosmologies
like the FLRW will \textit{not} produce observationally constant
number densities at moderate or high redshift values because this is
theoretically prohibited \citep[Sec.\ 2.1]{gabriela}. As discussed at
length elsewhere \citep{juracy2008}, observational and spatial
homogeneities are different relativistic concepts in cosmology, thus,
even a cosmological-principle-obeying spatially homogeneous cosmological
model will exhibit observational inhomogeneities at moderate and high
redshift ranges \citep[see also][]{ribeiro92b,ribeiro94,ribeiro95,
ribeiro2001b,ribeiro2005}. Therefore, $\gobs^\ast$ is an average
relativistic density and \textit{must not be confused with} the fluid
approximation local density $\rho$ appearing on the right hand side of
the Einstein equations.

Second, theoretical calculations of $\gobs^\ast$ along the past light
cone in the FLRW cosmologies had already predicted that a decay of
$\gobs^\ast$ at increasing observation distances is to be
observationally expected \citetext{\citealp{ribeiro92b}, Fig.\ 1;
\citealp{ribeiro95}, Figs.\ 1, 3; \citealp{ribeiro2001b}, Fig.\ 2},
which means that dealing with observations even in FLRW, or FLRW like,
cosmology backgrounds should lead to a decrease for $D$ at $z>1$. 

The third important relativistic concept that one must bear in mind
when dealing with relativistic fractal cosmologies is that number
densities in fractal cosmology are defined in terms of observational
distances, which means that at high redshift $\dobs$ will have
different values for each distance definition at the same redshift
value $z$. This means that as distance in relativistic cosmology is
not uniquely defined \citep{ellis71,ellis2007,holanda2010,holanda2011,
holanda2012} $\dobs$ must be replaced by $d_i$ in the equations above.
The index indicates the observed distance measure chosen to be
calculated with a specific redshift value. The distance definitions
applicable to relativistic fractal cosmology are the \textit{luminosity
distance} $\dl$, \textit{redshift distance} $\dz$, and \textit{galaxy
area distance} $\dg$, also known as \textit{transverse comoving
distance}. Two of these distance definitions are connected by the
Etherington reciprocity law below \citep{etherington33,ellis2007}, 
\be
\dl=(1+z)\,\dg.
\lb{eth}
\ee
The redshift distance is defined as, 
\be 
\dz=\frac{c \, z}{H_0},
\lb{red}
\ee
where $c$ is the light speed and $H_0$ is the Hubble constant.

Fractal analyses can be performed using $\dl$ and $\dg$ in any
cosmological model, however Eq.\ (\ref{red}) is only valid in FLRW
cosmologies. Besides, some caution is necessary with $\dg$ because
its respective volume density in the Einstein-de Sitter cosmology
results in $\gamg^\ast=\mbox{constant}$ \citep[pp.\ 1718, 1723-1724]
{ribeiro2001b}, although this is not true in all FLRW models, because
\citet[Fig.\ 7]{vinicius2007} and \citet[Figs.\ 1-12]{iribarrem2012a}
showed that in the $\Lambda$CDM cosmology the number densities
obtained with these three relativistic distances possess decaying
power-law properties. This conclusion justifies their adoption in the
present study. In addition, although $D$ can be
calculated with any distance, the comoving distance could be seen as
more appropriate because this is the distance where one often assumes
homogeneity to be present when one projects galaxies and fluctuations
in cosmology.

Bearing these points in mind, the expressions above must be
rewritten as below in order to become applicable to relativistic
cosmologies: 
\be
\dobs=d_i,
\lb{dists}
\ee
\be
\Vobs=V_i=\frac{4}{3} \pi {(d_i)}^3,
\lb{vi}
\ee
\be
\Nobs=N_i=B_i \, {(d_i)}^{D_i}, 
\lb{Nobs_i}
\ee
\be
\gobs^\ast=\gamma^\ast_i =\frac{N_i}{V_i}=\frac{3B_i}{4\pi}
{(d_i)}^{D_i-3},
\lb{gobs-ast_i}
\ee
where $i=({\ssty L}$, ${\sty z}$, ${\ssty G})$ according to the
chosen distance definition. The constant $B_i$ becomes attached
to each specific distance, the same being true for the fractal
dimension $D_i$. This is so because $N_i$ is counted considering
the limits given by each distance definition, which means that for
a given $z$ each $d_i$ will produce its respective $V_i$, $N_i$,
$B_i$ and $D_i$. Hence, all quantities become attached to a certain
distance definition

As final comments, one must stress again the difference between
spatial and observation number densities, difference which arises
only when the relativistic concept of past light cone, the
geometrical locus of astronomical observations, is taken into
account when modeling fractal cosmology at moderate and high
redshift scales. So, only by correctly manipulating the theoretical
tools of relativistic cosmological models that the possible
large-scale fractality of galaxy distribution will be revealed
\citep{ribeiro95, ribeiro2001b,ribeiro2005,juracy2008}.

\section{Fractal analysis}\lb{fractal-analysis}

Testing the fractal galaxy distribution hypothesis was done with
galaxy datasets provided by the COSMOS2015 and SPLASH redshift
surveys. Both catalogs contain hundreds of thousands additional
galaxies with measured redshifts as compared to the UltraVISTA DR1
dataset studied by \citet{teles2020}. Details on these surveys
relevant to the present study, followed by the fractal analyses
performed in their respective datasets, are shown below.

\subsection{The COSMOS2015 Galaxy Redshift Survey}\lb{DR2}

The COSMOS2015 catalog \citep{Laigle2016} includes
YJHK$_{\mathrm{S}}$ images from the UltraVISTA DR2, Y-band
images from Subaru/Hyper-Suprime-Cam and infrared data from
the Spitzer Large Area survey over 2 deg$^2$ in the COSMOS
field \citep{cosmos2007}. The object detection is performed
by the $\chi^2$ sum of the YJHK$_{\mathrm{S}}$ and z$^{++}$
images. Based on these data the UltraVISTA DR2 region contains
578379 galaxies, which means that 359079 new objects were added
to the sample as compared to the 219300 galaxies comprising the
UltraVISTA DR1. The observed area has the following range of
coordinates: $1.61\le\mbox{Dec (deg)}\le 2.81$ and $149.31
\le \mbox{Ra (deg)}\le150.79$, in which the full region has a
limiting magnitude K$_{\mathrm S}=24.0$ at $3\sigma$ in a 3"
diameter aperture and parts of the field covered by the
``ultra-deep stripes'' (0.62 deg$^2$) have limiting magnitude
K$_{\mathrm S}=24.7$ at $3\sigma$ in a 3" diameter. The
photometric redshifts were computed using LePHARE
\citep{arnouts,refId0} following \cite{ultra11}. The total
sample has $0.1\le z\le6$, with a very sizable number of
galaxies located in the scale of $1<z<4$ (see Fig.\ \ref{histz}).
Hence, these features placed the additional galaxies provided by
this survey well within the purposes of this study.

\subsection{The SPLASH Galaxy Redshift Survey}\lb{splash}

\begin{figure}
  \includegraphics[trim=2cm 0 0 0,scale=0.33]{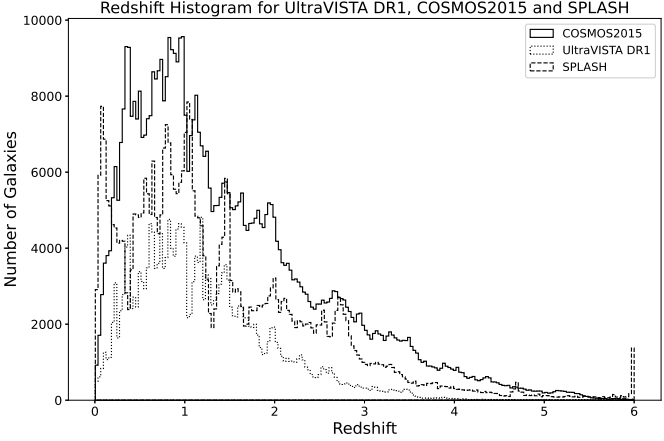}
  \caption{Histogram showing the galaxy distribution numbers in terms
  of redshift for the SPLASH, UltraVISTA DR1 and COSMOS2015
  surveys.}
  \label{histz}
\end{figure}

The SPLASH survey is a deep field galaxy redshift catalog with
multi-wavelength photometry within $2.4\,\mbox{deg}^{\,2}$ in the
sky region of $-5.64\le\mbox{Dec (deg)}\le-4.35$ and $33.84\le
\mbox{Ra (deg)}\le35.16$ \citep{Mehta2018}. The sources were
identified using a detection image defined as a $\chi^2$ combination
of grzy images from Hyper-Suprime-Cam (HSC) DR1, JHK images from
Ultra Deep Survey (UDS) DR11, YJHK$_{\mathrm S}$ images from VISTA
Deep Extragalactic Observations (VIDEO), u image from Megacam
Ultra-deep Survey: U-Band Imaging (MUSUBI), and ugri images from
CFHT Legacy Survey (CFHTLS). This catalog contains 390362 galaxies
at $0<z<6$, where the redshifts were measured using LePHARE with a
similar approach to that used for the COSMOS field \citep{ultra11,
Laigle2016}. The galaxy number distribution in terms of the redshift
is shown in Fig.\ \ref{histz}, where it is clear that a sizable
portion of its galaxies are in the redshift interval $1<z<4$, a fact
that justifies its inclusion in this fractal analysis, Besides, the
SPLASH galaxies were mapped in a different sky region as compared to
the surveys discussed above, and it follows a different strategy for
the detection of the objects, as it includes the u-band, in order to
recover the bluest objects.

\subsection{Data Filtering}

Fractal analysis of galaxy surveys requires disposing the data along
volume-limited distributions. As galaxy surveys are limited by apparent
magnitude, we must proceed by reducing the data into subsamples such
that they follow increasing redshift bins. This is done by plotting
the absolute magnitudes of galaxies in terms of their respective
measured redshifts, and then by only choosing galaxies below a certain
absolute magnitude threshold defined by the limiting apparent magnitude
of the survey. The usual expression
\be
M=m-5\log \dl (z) -25,
\lb{magabs1}
\ee
where $M$ is the absolute magnitude, $m$ is the apparent magnitude and
$\dl$ is given in Mpc, can be used for this purpose. We assumed the FLRW
cosmology with $\Omega_{m_{\ssty 0}}=0.3$, $\Omega_{\Lambda_{\ssty 0}}=0.7$
and $H_0=(70\pm5)\;\mbox{km}\; {\mbox{s}}^{-1} \; {\mbox{Mpc}}^{-1}$.

Next, the apparent magnitude threshold in the K-band was assumed to be
${\mathrm K}=24.7$, a mean limiting value acceptable to both surveys
\cite[see][]{teles2020} such that homogeneous volume-limited subsamples
could be created by changing Eq.\ (\ref{magabs1}) to the expression below,
\be
M_{\mathrm K}=24.7-5\log \dl (z) -25,
\lb{magabs2}
\ee
which provides the cutoff line between the filtered
and unfiltered galaxies. Finally, this whole process
was made effective considering the above uncertainty in the Hubble
constant in order to test to what extent, if any, our results would be
affected by its error margin.

\begin{figure*}
  \includegraphics[scale=1.11]{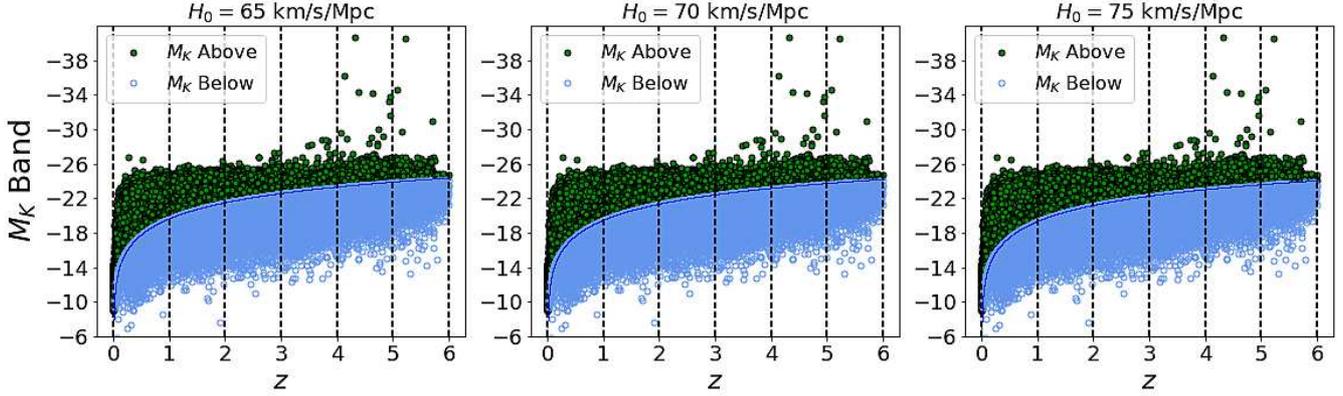}
  \caption{Plot of the absolute magnitudes for the COSMOS2015
  galaxies in terms of of their photometrically measured redshift
  values. The dividing line corresponds to apparent magnitude
  ${\mathrm K}=24.7$ and only galaxies having $M_{\mathrm K}$ above
  this cutoff and $z\leq4$ were included in subsamples that assumed
  three different values of the Hubble constant.}
  \label{magcomplete-dr2}
\end{figure*}

\begin{figure*}
  \includegraphics[scale=1.11]{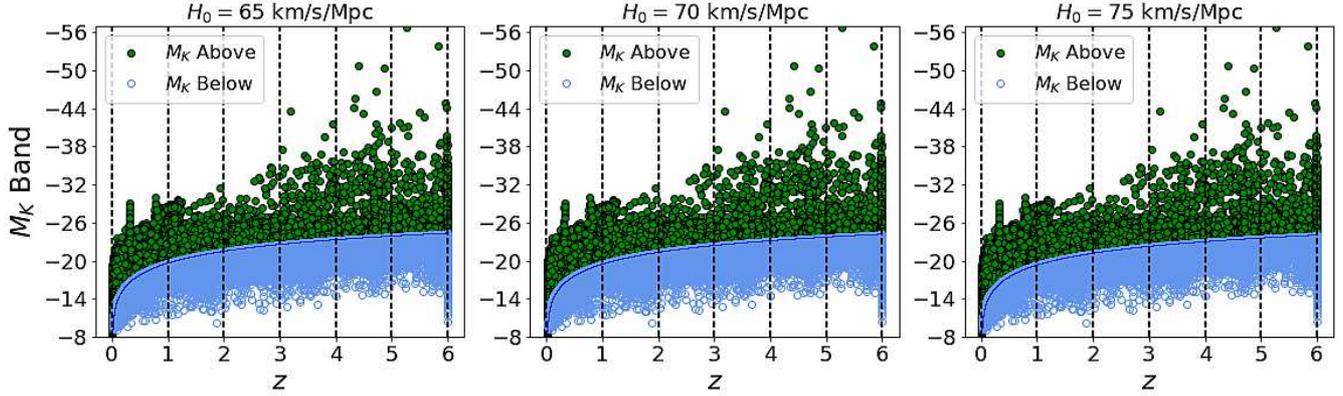}
  \caption{Plot of the absolute magnitudes for the SPLASH galaxies in
  terms of of their photometrically measured redshift values. The
  dividing line corresponds to apparent magnitude ${\mathrm K}=24.7$
  and only galaxies having $M_{\mathrm K}$ above this cutoff and
  $z\leq4$ were included in subsamples that assumed three different
  values of the Hubble constant within its uncertainty.}
  \label{magcomplete-splash}
\end{figure*}

Fig.\ \ref{magcomplete-dr2} shows the selection of the COSMOS2015
survey obtained according to the filtering procedure
described above. Only galaxies with absolute magnitudes $M_{\mathrm K}$
above the cutoff line were included in a subsample for further analysis.
In addition, we also disregarded galaxies having $z>4$ as, according to
Fig.\ \ref{histz}, their numbers are too small to be considered
representative. The end result of this process was the creation of
three subsamples containing 230705 galaxies in the redshift range
$0.004\leq z\leq4$ out of the original 578379 objects for three values
of the Hubble constant within its uncertainty.
Fig.\ \ref{magcomplete-splash} shows the same filtering procedure
carried out with the SPLASH galaxies generating three other subsamples
as well. The original 390362 objects were then reduced to 171548 galaxies.

\subsection{Data Analysis}\lb{data-analysis}

The above parameters for the FLRW cosmology allowed the calculation
of the observational distances $d_i\,(i={\ssty G}$, ${\ssty L}$,
${\sty z})$ using theoretical expressions relating distance to redshift
provided by this cosmological model with the photometric redshift
values furnished by the galaxy surveys. The data sorting algorithm
necessary for carrying out a fractal analysis under the theoretical
model discussed in Sec.\ \ref{fractal-cos} above is the same as
employed by \citet{teles2020}. It is suscintly described as follows. 

We started by establishing the minimum redshift value $z_{\ssty 0}$,
the respective minimum distances $d_{i_0}=d_{i_0}(z_{\ssty 0})$, and
the incremental distance interval $\Delta d_i$. The algorithm was
initiated by counting the number of observed galaxies $N_{i_1}$
in the first interval $d_{i_1}=d_{i_0}+\Delta d_i$ and calculating
the respective volume density $\gamma_{i_1}^\ast$. This defined the
first bin. The next step was to increase the bin size by $\Delta d_i$.
Values for $N_{i_2}$ and $\gamma_{i_2}^\ast$ were then calculated at
the distance $d_{i_2}=d_{i_0}+2\Delta d_i$. These steps were
repeated $n$ times until the farthest group of galaxies were included
and all quantities of interest counted and calculated.

We tested different bin size increments $\Delta d_i$ for each
distance definition to see if the results would be affected, with
negative results. Therefore, the interval $\Delta d_i=200$~Mpc
was applied to all calculations, choice which in the end provided
a large amount of data points for all quantities involved, allowing
then enough points to perform adequate regression analyses.

Finally, according to Eq.\ (\ref{gobs-ast_i}) plots of $\gamma_i^\ast$
vs.\ $d_i$ would behave as decaying power-law curves if the galaxy
distribution really formed a fractal system. In this case the linear
fit slopes in log-log plots would allow for the fractal dimensions
$D_i$ of the distribution to be directly determined.

\section{Results}\lb{results}

\begin{figure*}
  \includegraphics[scale=0.94]{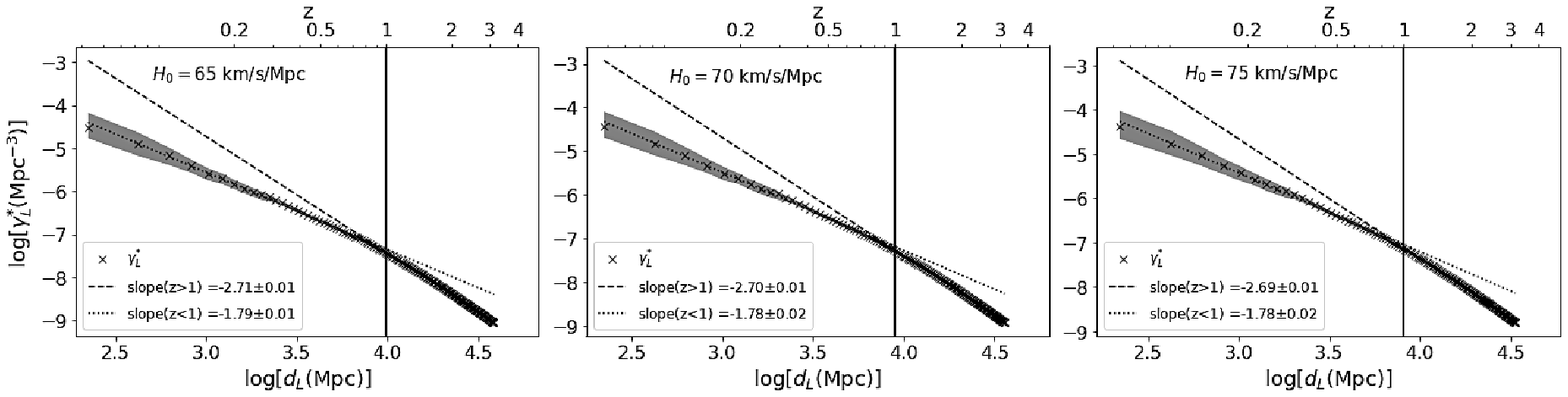}
  \caption{Log-log graph of $\gaml^\ast$ vs.\ $\dl$ obtained with
           the COSMOS2015 galaxy redshift survey dataset in the
           ranges $z<1$ and $1\leq z\leq4$ and respective distance
           measures.} 
  \label{gammaLdL-dr2}
\end{figure*}

\begin{figure*}
 \includegraphics[scale=0.94]{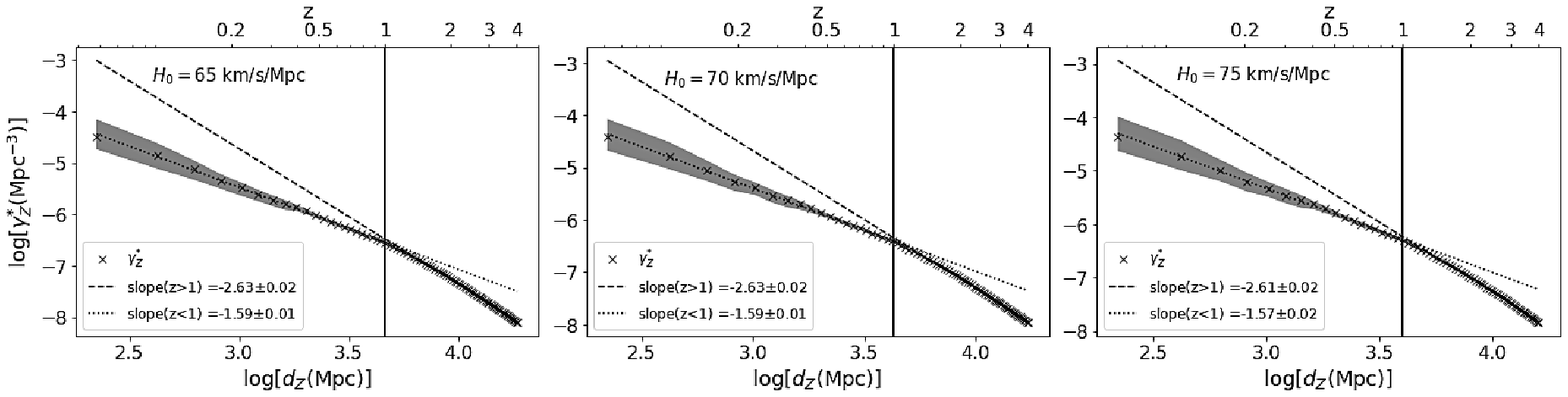}
  \caption{Log-log graph of $\gamz^\ast$ vs.\ $\dz$ obtained with
           the COSMOS2015 galaxy redshift survey dataset in the
           ranges $z<1$ and $1\leq z\leq4$ and respective distance
           measures.} 
  \label{gammaZdZ-dr2}
\end{figure*}

\begin{figure*}
 \includegraphics[scale=0.94]{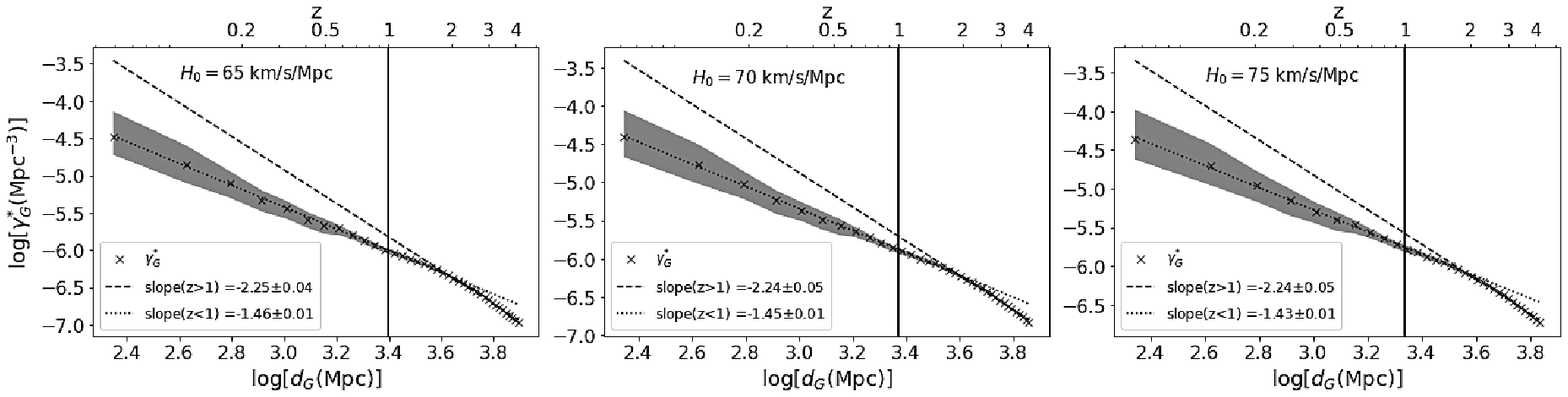}
  \caption{Log-log graph of $\gamg^\ast$ vs.\ $\dg$ obtained with
           the COSMOS2015 galaxy redshift survey dataset in the
           ranges $z<1$ and $1\leq z\leq4$ and respective distance
           measures.} 
  \label{gammaGdG-dr2}
\end{figure*}

\begin{figure*}
 \includegraphics[scale=0.94]{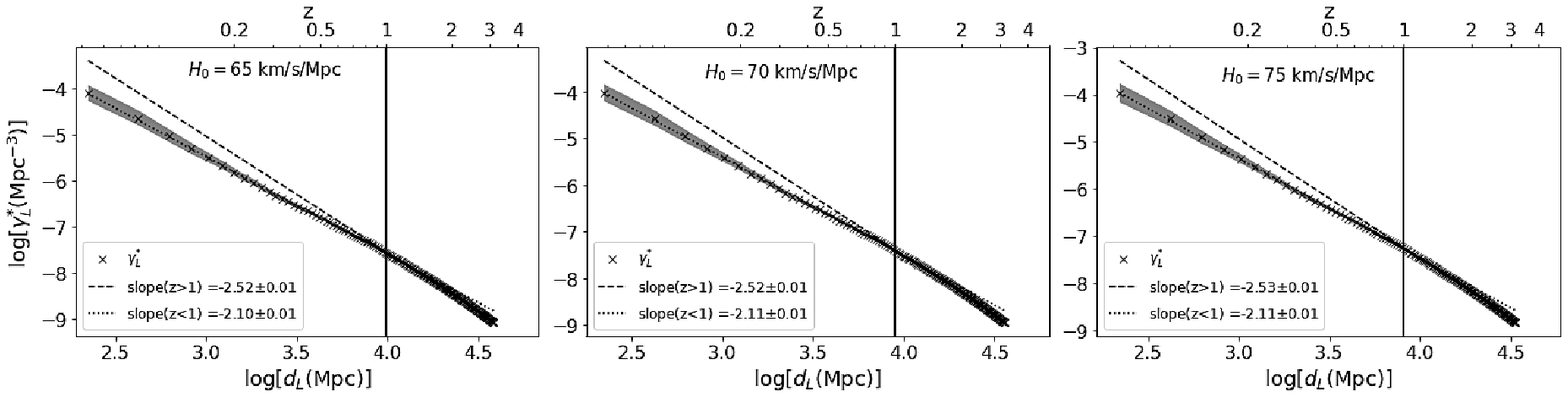}
  \caption{Log-log graph of $\gaml^\ast$ vs.\ $\dl$ obtained with
           the SPLASH galaxy redshift survey dataset in the ranges
           $z<1$ and $1\leq z\leq4$ and respective distance measures.} 
  \label{gammaLdL-splash}
\end{figure*}

\begin{figure*}
 \includegraphics[scale=0.94]{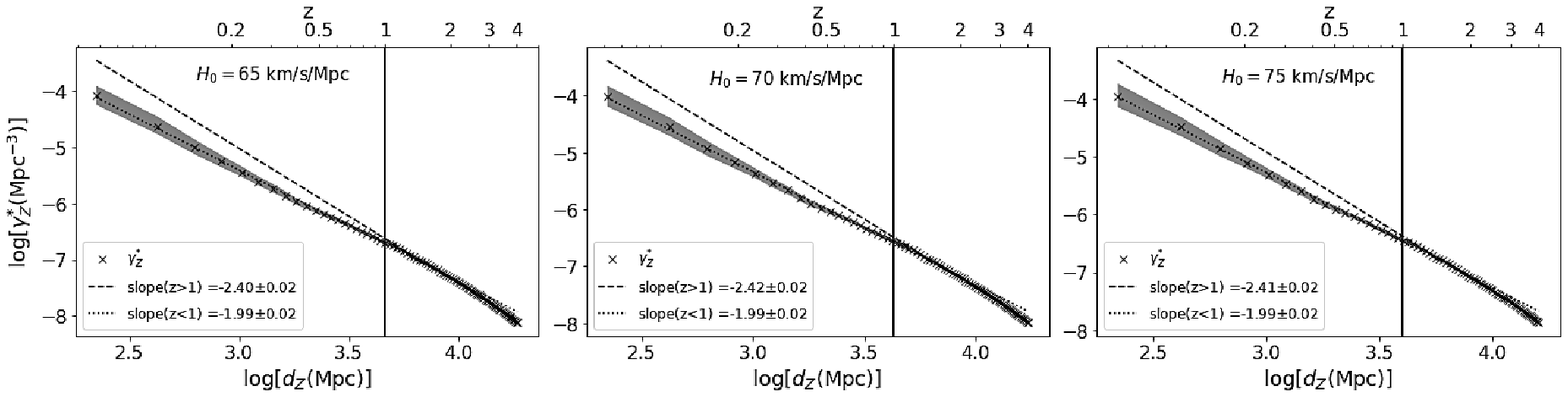}
  \caption{Log-log graph of $\gamz^\ast$ vs.\ $\dz$ obtained with
           the SPLASH galaxy redshift survey dataset in the ranges
           $z<1$ and $1\leq z\leq4$ and respective distance measures.} 
  \label{gammaZdZ-splash}
\end{figure*}

\begin{figure*}
 \includegraphics[scale=0.94]{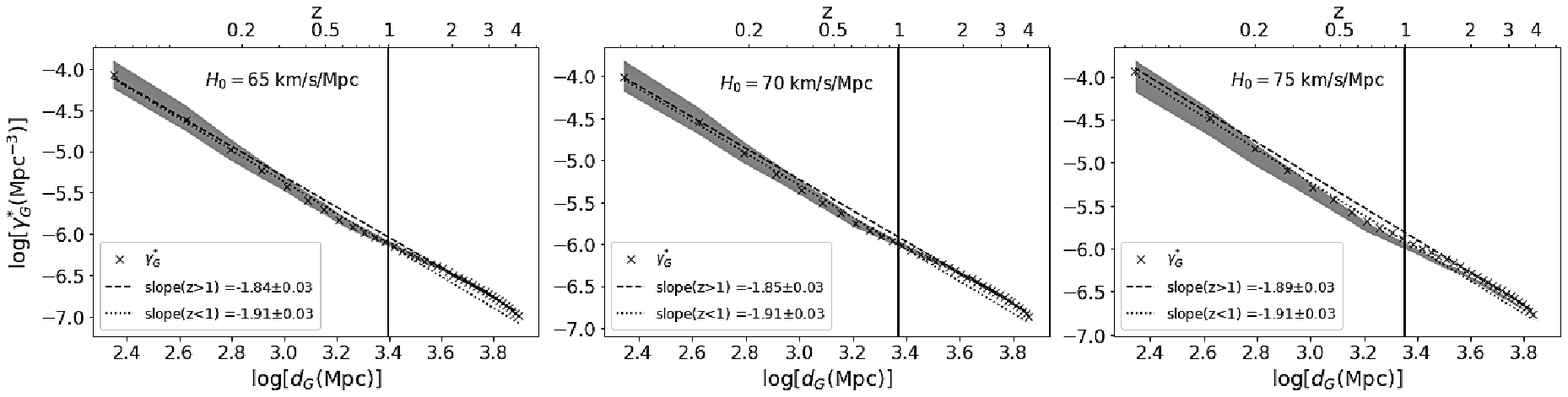}
  \caption{Log-log graph of $\gamg^\ast$ vs.\ $\dg$ obtained with
           the SPLASH galaxy redshift survey dataset in the ranges
           $z<1$ and $1\leq z\leq4$ and respective distance measures.} 
  \label{gammaGdG-splash}
\end{figure*}

Figs.\ \ref{gammaLdL-dr2} to \ref{gammaGdG-splash} show log-log graphs
of $\gamma_i^\ast$ vs.\ $d_i$ with both surveys' datasets studied here
with their respective choices of Hubble constant values within the above
defined uncertainty. The results show that the galaxy distribution in
both surveys present power-law decays in two scale ranges: for $z<1$ and
$1\leq z\leq 4$. This is consistent with a fractal system possessing two
single fractal dimensions at different distance ranges.

The fractal dimensions in both scaling ranges can be simply
calculated from the slopes of the fitted straight lines by means
of Eq.\ (\ref{gobs-ast_i}). All obtained results are collected in
Tables \ref{tab1} to \ref{tab4}, where one can clearly verify two 
single fractal systems in sequence at different data ranges
with decreasing values for $D$ at higher redshift ranges, as
theoretically predicted (see Sec.\ \ref{intro} above). Besides,
the results show the fractal dimensions being unaffected by
variations in the Hubble constant. 

One can summarize all results of Tables \ref{tab1} to \ref{tab4}
by conservatively rounding off the main results and uncertainties
around their medians. Hence, for $z<1$ the COSMOS2015 survey
produced $D=1.4\pm0.2$, whereas the SPLASH galaxies yielded
$D=1.0\pm0.1$. For $1\leq z\leq4$ we respectively found
$D=0.5\pm0.3$ and $D=0.8\pm0.4$. Clearly the SPLASH galaxies
produce fractal dimensions somewhat smaller than the COSMOS2015
ones for $z<1$, but the reverse situation for $z>1$, although with
overlapping uncertainties. Possible reasons for such differences and
comparison with previous studies will be discussed below.
\begin{table}
\caption{Fractal dimensions calculated in the reduced and
	volume-limited subsamples of the COSMOS2015
	galaxy redshift survey in the range $z<1$. The single
	fractal dimensions $D_{\ssty L}$, $D_{\ssty z}$ and
	$D_{\ssty G}$ were obtained from the galaxy distributions
	respectively using the luminosity distance $\dl$, redshift
	distance $\dz$ and galaxy area distance (transverse
        comoving distance) $\dg$.}
\label{tab1}
\begin{center}
\begin{tabular}{cccc}
\hline
$z<1$ & $D_{\ssty L}$ & $D_{\ssty z}$ & $D_{\ssty G}$\\
\hline
$H_0=65$ km/s/Mpc & $1.21\pm0.01$ & $1.41\pm0.01$ & $1.54\pm0.01$\\
$H_0=70$ km/s/Mpc & $1.22\pm0.02$ & $1.41\pm0.01$ & $1.55\pm0.01$\\
$H_0=75$ km/s/Mpc & $1.22\pm0.02$ & $1.43\pm0.02$ & $1.57\pm0.01$\\
\hline
\end{tabular}
\end{center}
\end{table}
\begin{table}
\caption{Fractal dimensions obtained with the COSMOS2015 survey
         subsamples in the range $1\leq z\leq4$. Quantities are as
         described in Table \ref{tab1}.} 
\label{tab2}
\begin{center}
\begin{tabular}{cccc}
\hline
$1\leq z\leq4$ & $D_{\ssty L}$ & $D_{\ssty z}$ & $D_{\ssty G}$\\
\hline
$H_0=65$ km/s/Mpc & $0.29\pm0.01$ & $0.37\pm0.02$ & $0.75\pm0.04$\\
$H_0=70$ km/s/Mpc & $0.30\pm0.01$ & $0.37\pm0.02$ & $0.76\pm0.05$\\
$H_0=75$ km/s/Mpc & $0.31\pm0.01$ & $0.39\pm0.02$ & $0.76\pm0.05$\\
\hline
\end{tabular}
\end{center}
\end{table}
\begin{table}
\caption{Fractal dimensions obtained with the SPLASH survey
         subsamples in the range $z<1$. Quantities are as
         described in Table \ref{tab1}.} 
\label{tab3}
\begin{center}
\begin{tabular}{cccc}
\hline
$z<1$ & $D_{\ssty L}$ & $D_{\ssty z}$ & $D_{\ssty G}$\\
\hline
$H_0=65$ km/s/Mpc & $0.90\pm0.01$ & $1.01\pm0.02$ & $1.09\pm0.03$\\
$H_0=70$ km/s/Mpc & $0.89\pm0.01$ & $1.01\pm0.02$ & $1.09\pm0.03$\\
$H_0=75$ km/s/Mpc & $0.89\pm0.01$ & $1.01\pm0.02$ & $1.09\pm0.03$\\
\hline
\end{tabular}
\end{center}
\end{table}
\begin{table}
\caption{Fractal dimensions obtained with the SPLASH survey
         subsamples in the range $1\leq z\leq4$. Quantities are as
         described in Table \ref{tab1}.} 
\label{tab4}
\begin{center}
\begin{tabular}{cccc}
\hline
$1\leq z\leq4$ & $D_{\ssty L}$ & $D_{\ssty z}$ & $D_{\ssty G}$\\
\hline
$H_0=65$ km/s/Mpc & $0.48\pm0.01$ & $0.60\pm0.02$ & $1.16\pm0.03$\\
$H_0=70$ km/s/Mpc & $0.48\pm0.01$ & $0.58\pm0.02$ & $1.15\pm0.03$\\
$H_0=75$ km/s/Mpc & $0.47\pm0.01$ & $0.59\pm0.02$ & $1.11\pm0.03$\\
\hline
\end{tabular}
\end{center}
\end{table}

Finally, for comparison of fractal dimensions obtained with similar
methodology as described here, Table \ref{tab5} presents all values for
$D$ calculated with the UltraVISTA DR1, COSMO2015, SPLASH and FDF
surveys when assuming $H_0=70$ km/s/Mpc.
\begin{table*}
\caption{This table presents a comparison of all recently calculated single
	fractal dimensions applying similar analytical tools as presented
	here to various galaxy distribution surveys. These results were
	obtained with the UltraVISTA DR1 \protect\citep{teles2020},
	COSMOS2015 and SPLASH (this work), and FDF \protect\citep{gabriela}
	catalogs, all considering $H_0=70$ km/s/Mpc. There is a clear
	tendency for decreasing values of $D$ at $z>1$ in virtually all
	results, as theoretically predicted (see Sec.\ \ref{intro} above).
	Such a decrease is, nonetheless, less pronounced in the SPLASH data,
	which is the only galaxy distribution shown here to have been
	surveyed in the southern hemisphere.}
\label{tab5}
\begin{center}
\begin{tabular}{ccccccccc}
\hline
& UVista DR1 & $(0.2<z<4)$ & COSMO2015 & $(0.1<z<4)$ &
SPLASH & $(0.1<z<4)$ & FDF & $(0.45<z<5)$ \\
\hline
 & $z<1.0$ & $z>1.0$ & $z<1.0$ & $z>1.0$ & $z<1.0$ & $z>1.0$ &
 $z\lesssim1.2$ & $z\gtrsim1.2$ \\
\hline
$D_{\ssty L}$ & $1.40\pm0.02$ & $0.32\pm0.01$ & $1.22\pm0.02$ & $0.30\pm0.01$ &
   $0.89\pm0.01$ & $0.48\pm0.01$ & $1.2\pm0.3$ & $0.5\pm0.2$ \\
\hline
$D_{\ssty z}$ & $1.61\pm0.02$ & $0.38\pm0.02$ & $1.41\pm0.01$ & $0.37\pm0.02$ &
   $1.01\pm0.02$ & $0.58\pm0.02$ & $1.5\pm0.4$ & $0.6\pm0.2$ \\
\hline
$D_{\ssty G}$ & $1.75\pm0.03$ & $0.81\pm0.06$ & $1.55\pm0.01$ & $0.76\pm0.05$ &
   $1.09\pm0.03$ & $1.15\pm0.03$ & $1.8\pm0.3$ & $1.0\pm0.7$ \\
\hline
\end{tabular}
\end{center}
\end{table*}

\section{Blue and Red subsamples}\lb{blue-red}

All single fractal dimensions presented so far were calculated without
any consideration of galactic types, features or possible evolutionary
stages. Hence, one may wonder if $D$ could depend on some, or all, of
these characteristics, however they are defined or observed. If such
possible dependencies are actually found, fractal dimensions could,
perhaps, be used as tracers of galactic properties or their
evolutionary stages. Below we propose a simple test of this possible
dependency using the surveys studied here.

The COSMOS2015 and SPLASH data allow us to calculate $D$ in two galaxy
subsamples: the blue, or star forming, and the red, or quiescent,
galaxies. Such a selection provides a preliminary and straightforward
way of testing the concept of possible use of $D$ as a tracer of
galactic features.\footnote{We are grateful to a referee for suggesting
this test.} The criteria for generating these subsamples use color-color
diagrams or star formation rates as provided in both surveys databases.

For the COSMOS2015 dataset, the classification presented in
\cite{Laigle2016} is derived from the location of galaxies in the
NUV-r vs.\ r-J color-color diagram \citep{Wi2009}. Besides using these
colors, this selection has the estimation of the absolute magnitudes at
rest-frame based on the apparent magnitudes at
$\lambda_{\mbox{\footnotesize rest-frame}}(1+z_{\mbox{\footnotesize
gal}})$, which minimises the k-correction dependency \citep{Il2005}.
Such technique avoids the mixing between red dusty galaxies and
quiescent ones. In practice, included quiescent galaxies have
${M}_{\mathrm{\ssty NUV}}-{M}_{\mathrm r}> 3\,({M}_{\mathrm
r}-{M}_{\mathrm J})+1$ and ${M}_{\mathrm{\ssty NUV}}-{M}_{\mathrm
r}> 3.1$, while the others were considered star-forming.

Regarding the SPLASH survey, following \cite{Il2010} and
\cite{DS2011} we separated the galaxy sample using the \textit{specific
star formation rate} (SSFR). Galaxies with $\log \mbox{SSFR}<-11$ were 
classified as quiescent, whereas the ones with $\log \mbox{SSFR}>-11$
were set as star-forming. SSFR is the ratio between star formation
rate and stellar mass, being a simple way of quantifying in a uniform
way the degree of star formation. Such division by stellar mass gives
an indication of how strong is the star formation in a certain galaxy,
this being especially important when comparing galaxies having different
sizes and masses. In \cite{ultra11} such classification approach was 
compared to the color-color selection applied to COSMOS2015, and they
showed that both methods provide similar results at $z<1$, but towards
high redshift the classification based on the SSFR tends to be more
conservative.

\begin{figure*}
 \includegraphics[scale=0.90]{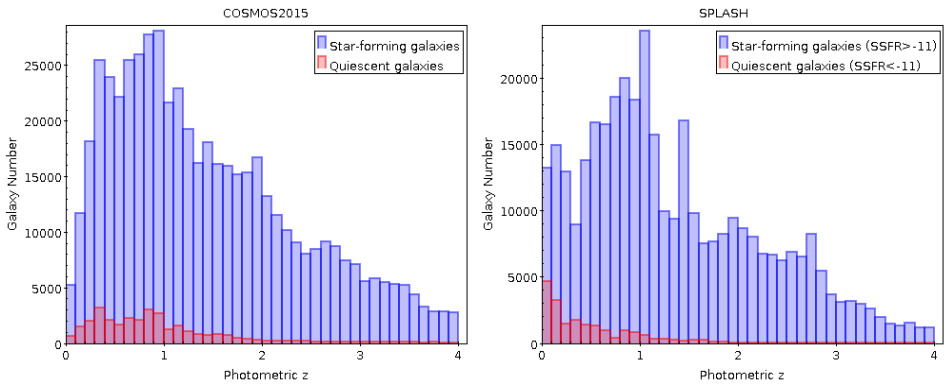}
 \caption{Histograms showing the galaxy distribution
          numbers in terms of the redshift for the COSMOS2015 (left) and
	  SPLASH (right) subsamples of blue star forming galaxies and red
	  quiescent ones. Labels are as in the legends. The COSMOS2015
	  galaxies were separated by color and SPLASH galaxies were
	  classified considering the \textit{specific stellar formation
	  rate} (SSFR) using -11 as the cutoff value (see the main text).
	  Clearly the number of blue galaxies is much higher than the red
	  ones in both surveys.} 
  \label{blue-red-cosmos-splash-classes}
\end{figure*}

\begin{figure*}
 \includegraphics[scale=0.50]{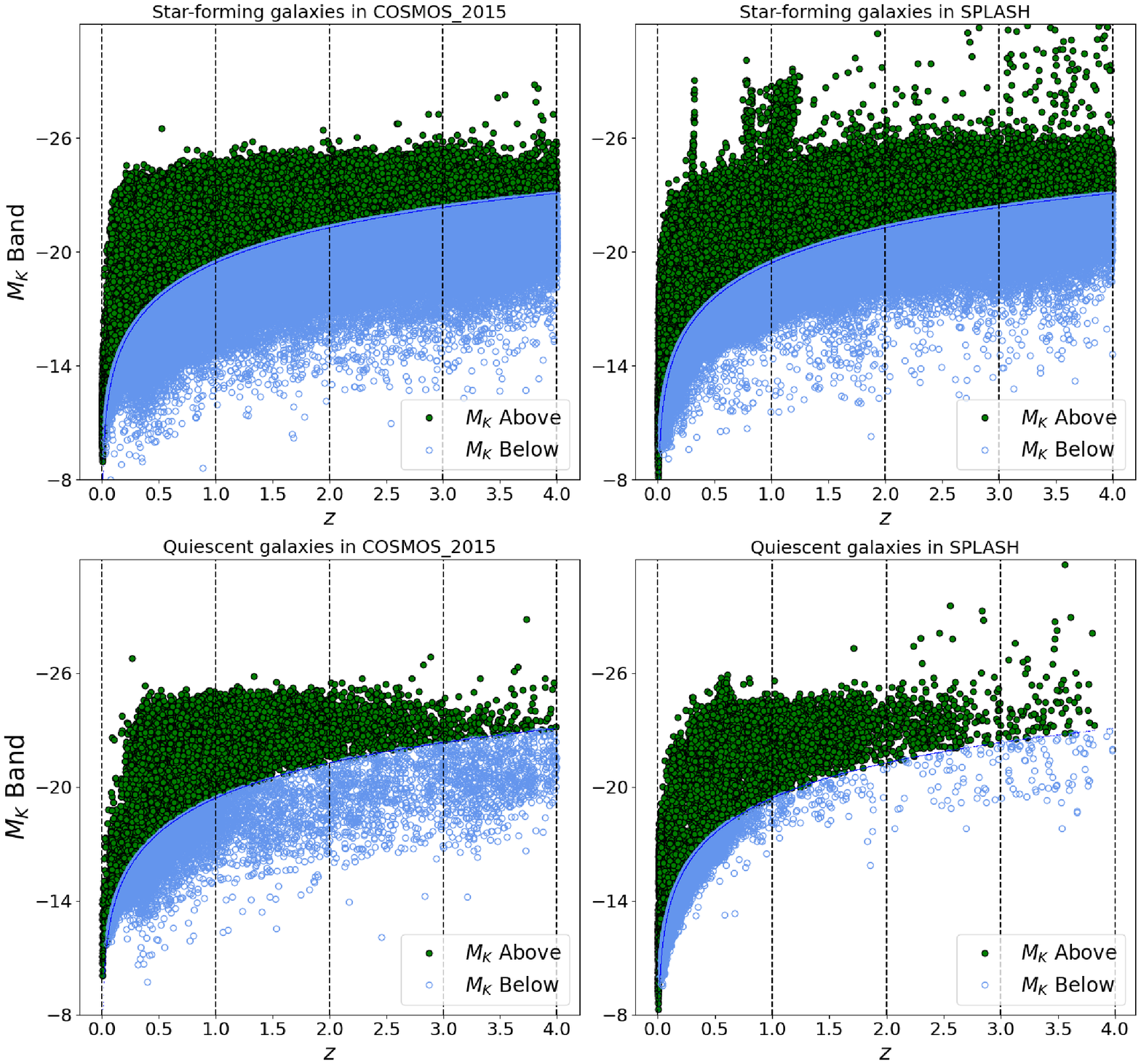}
 \caption{Plots of the absolute magnitudes for the blue
         star forming galaxies (top), and red quiescent ones (bottom)
	 for the COSMOS2015 (left) and SPLASH (right) surveys in terms
	 of their photometrically measured redshift values. Specific
	 discriminations are as in the title of each plot. As in Figs.\
	 \ref{magcomplete-dr2} and \ref{magcomplete-splash}, the dividing
	 line corresponds to apparent magnitude ${\mathrm K} = 24.7$, so
	 only blue and red galaxies having $M_{\mathrm K}$ above this
	 cutoff and $z\leq 4$ were included in the blue and red COSMOS2015
	 and SPLASH subsamples. Since previous results indicate that the
	 fractal dimension is not affected by varying the Hubble constant
	 within its currently accepted uncertainty, here only $H_0=70$
	 km/s/Mpc was assumed.} 
  \label{mag_complete-blue-red-cosmos-splash}
\end{figure*}

Considering the criteria above two subsamples were generated for each
survey. Fig.\ \ref{blue-red-cosmos-splash-classes} shows histograms
of redshift distribution of both subsamples up to $z=4$, where it is
clear that the number of blue star forming galaxies is much higher
than the red quiescent ones. This selection led the COSMOS2015 survey
ending up with 527899 star forming galaxies and 31424 quiescent ones.
The SPLASH had respectively 359021 and 20045 galaxies. These four
subsamples were then subjected to the same volume-limited filtering
process of absolute magnitude cutoff as defined by Eq.\ (\ref{magabs2}).
Fig.\ \ref{mag_complete-blue-red-cosmos-splash} shows the outcome of
this filtering procedure, which in the end left the COSMOS2015
subsamples further reduced to 208005 blue galaxies and 22824 red ones,
whereas the filtered SPLASH catalog subsamples respectively ended up
with 205012 and 13491 galaxies.

\begin{figure*}
 \includegraphics[scale=0.39]{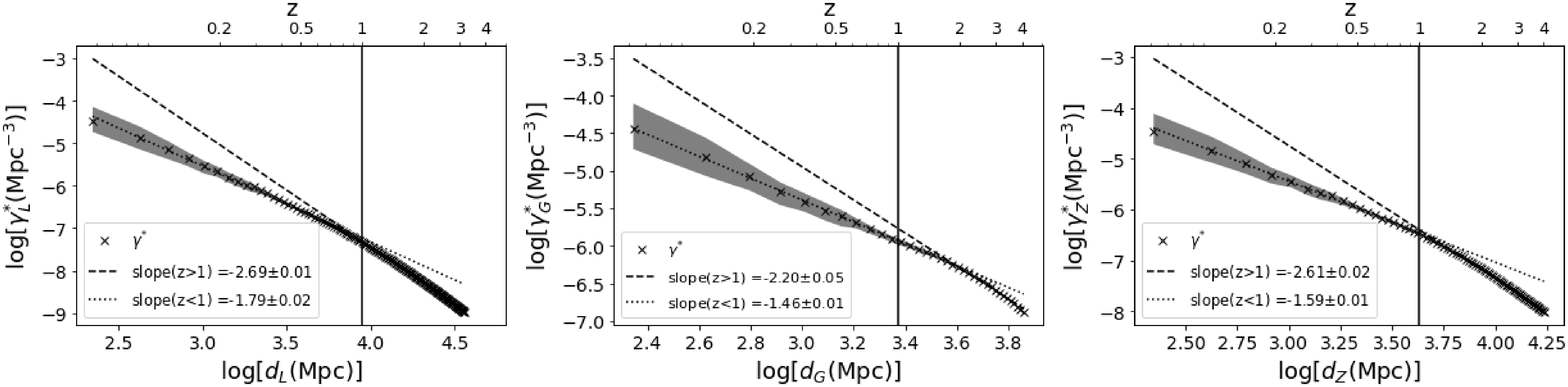}
 \caption{Log-log graph of $\gaml^\ast$, $\gamg^\ast$
          and $\gamz^\ast$ respectively vs.\ $\dl$, $\dg$ and $\dz$
	  obtained with the blue COSMOS2015 star forming galaxy subsample
	  in the ranges $z<1$ and $1\leq z\leq 4$ and respective distance
	  measures. All results were obtained considering $H_0=70$ km/s/Mpc.}
\label{bluecosmosdLdGdz}
\end{figure*}

\begin{figure*}
 \includegraphics[scale=0.39]{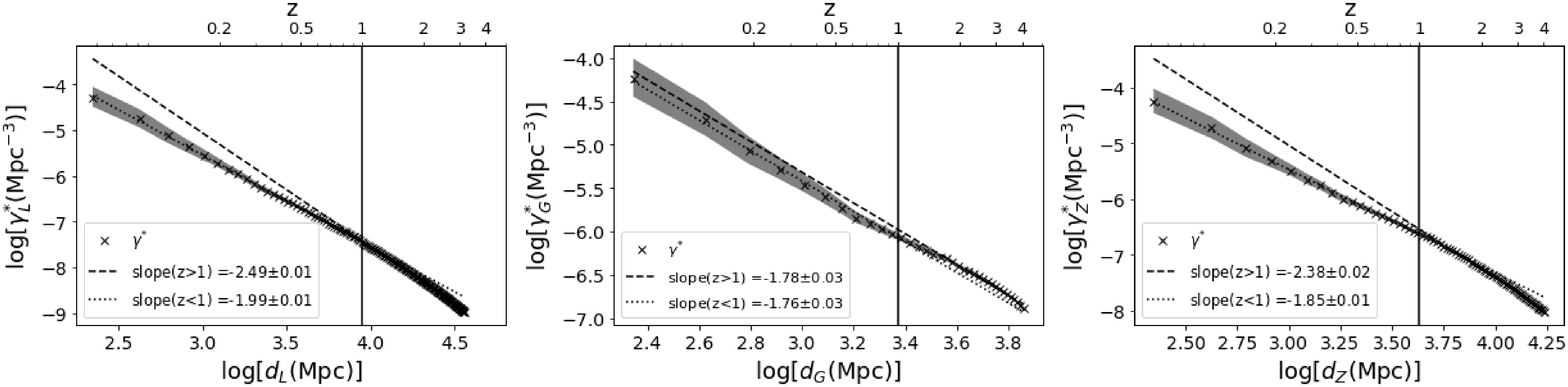}
 \caption{Log-log graph of $\gaml^\ast$, $\gamg^\ast$
          and $\gamz^\ast$ respectively vs.\ $\dl$, $\dg$ and $\dz$
	  obtained with the blue SPLASH star forming galaxy subsample
	  in the ranges $z<1$ and $1\leq z\leq 4$ and respective distance
	  measures. All results were obtained considering $H_0=70$ km/s/Mpc.}
\label{bluesplashdLdGdz}
\end{figure*}

\begin{figure*}
\includegraphics[scale=0.385]{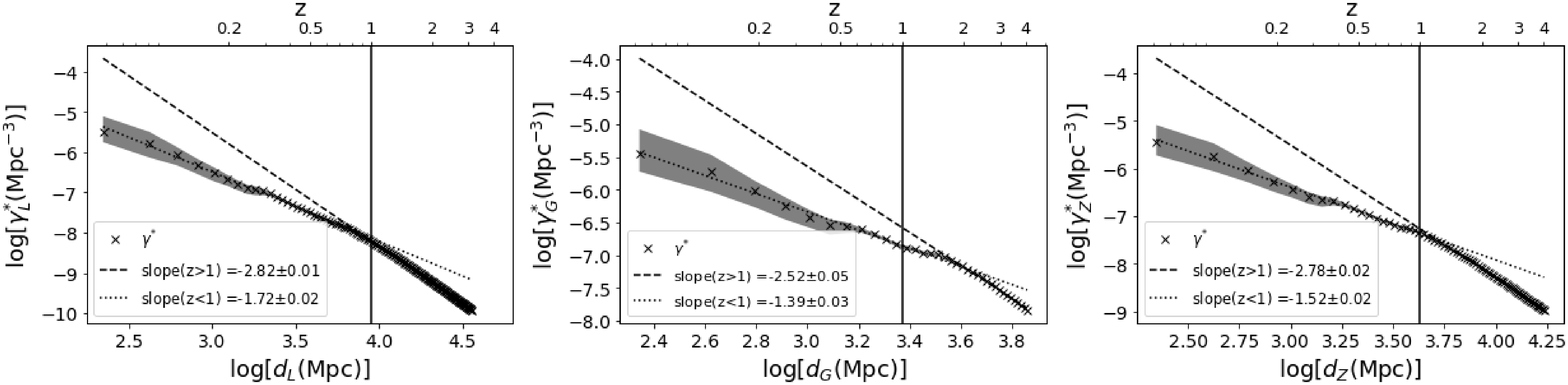}
\caption{Log-log graph of $\gaml^\ast$, $\gamg^\ast$
          and $\gamz^\ast$ respectively vs.\ $\dl$, $\dg$ and $\dz$
	  obtained with the red COSMOS2015 quiescent galaxy subsample
	  in the ranges $z<1$ and $1\leq z\leq 4$ and respective
	  distance measures. All results were obtained considering
	  $H_0=70$ km/s/Mpc.}
\label{redcosmosdLdGdz}
\end{figure*}

\begin{figure*}
\includegraphics[scale=0.39]{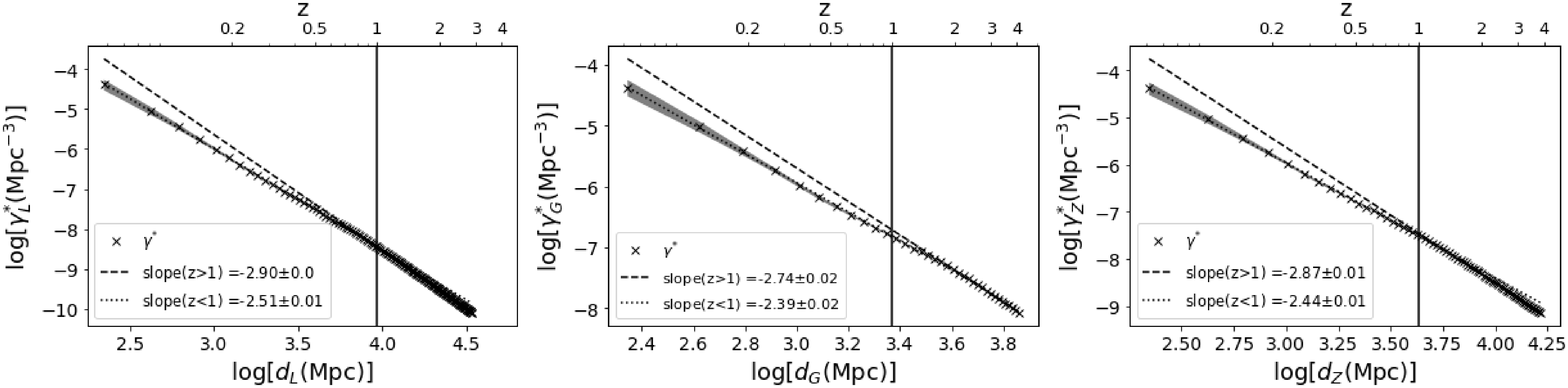}
\caption{Log-log graph of $\gaml^\ast$, $\gamg^\ast$
          and $\gamz^\ast$ respectively vs.\ $\dl$, $\dg$ and $\dz$
	  obtained with the red SPLASH quiescent galaxy subsample in
	  the ranges $z<1$ and $1\leq z\leq 4$ and respective distance
	  measures. All results were obtained considering $H_0=70$
	  km/s/Mpc.}
\label{redsplashdLdGdz}
\end{figure*}

The selected and filtered galaxies of the four subsamples had their
$\gamma_i^\ast$ number densities calculated using the same procedure
as described in Sec.\ \ref{data-analysis} above. The resulting data
points were then linear fitted against their respective distance
measures $d_i\,(i={\ssty G}$, ${\ssty L}$, ${\sty z})$. Figs.\
(\ref{bluecosmosdLdGdz}) to (\ref{redsplashdLdGdz}) show the decaying
power-law curves and actual data fits in the ranges $z<1$ and
$1\leq z\leq 4$. In view of the results presented in Tables \ref{tab1}
to \ref{tab4}, which showed robustness of the fractal dimension against
changes in the Hubble constant, in this Section we calculated all
results using only $H_0=70$ km/s/Mpc.
\begin{table*}
	\caption{Fractal dimensions calculated in the selected
	blue-star-forming and red-quiescent and then
	volume-limit-filtered galaxy subsamples of the COSMOS2015 and
	SPLASH redshift surveys in the range $z\leq 4$. The single
	fractal dimensions $D_{\ssty L}$, $D_{\ssty z}$ and $D_{\ssty G}$
	were obtained from the galaxy distributions respectively using
	the luminosity distance $\dl$, redshift distance $\dz$ and
	galaxy area distance (transverse comoving distance) $\dg$. The
	results were calculated considering only $H_0=70$ km/s/Mpc (see
	the main text). A comparison of these figures with the ones in the
	3rd to 6th columns of Table \ref{tab5} shows that the
	fractal dimensions do vary according to the blue-red selection
	used here, this being especially the case for the quiescent
	galaxies. In addition, as in the results shown in Table
	\ref{tab5}, all values of $D$ decrease in the range $z>1$,
	in some cases quite substantially.}
\label{tabA}
\begin{center}
\begin{tabular}{ccccccccc}
\hline
& Blue COSMOS2015 & $\sty (0.1<z<4)$ &
  Blue SPLASH & $\sty (0.1<z<4)$ &
  Red COSMOS2015 & \scalebox{0.8}{$(0.1<z<4)$} &
  Red SPLASH & $\sty (0.1<z<4) $ \\
\hline
& $z<1.0$ & $z>1.0$ & $z<1.0$ & $z>1.0$ & $z<1.0$ &
$z>1.0$ & $z<1.0$ & $z>1.0$ \\
\hline
$D_{\ssty L}$ & $1.21\pm0.02$ & $0.31\pm0.01$ & $1.01\pm0.01$ & $0.51\pm0.01$ &
   $1.28\pm0.02$ & $0.18\pm0.01$ & $0.49\pm0.01$ & $0.10\pm0.01$ \\
\hline
$D_{\ssty z}$ & $1.41\pm0.01$ & $0.39\pm0.02$ & $1.15\pm0.01$ & $0.62\pm0.02$ &
   $1.48\pm0.02$ & $0.22\pm0.02$ & $0.56\pm0.01$ & $0.13\pm0.01$ \\
\hline
$D_{\ssty G}$ & $1.54\pm0.01$ & $0.80\pm0.05$ & $1.24\pm0.03$ & $1.22\pm0.03$ &
   $1.61\pm0.03$ & $0.48\pm0.05$ & $0.61\pm0.02$ & $0.26\pm0.02$ \\
\hline
\end{tabular}
\end{center}
\end{table*}

The fractal dimensions obtained from the graphs in Figs.\
(\ref{bluecosmosdLdGdz}) to (\ref{redsplashdLdGdz}) are collected
in Table~\ref{tabA}, which provides results that can now be compared with
the 3rd to 6th columns of Table \ref{tab5}. The blue COSMOS2015 galaxies
had their fractal dimensions basically unchanged from the unselected
samples, whereas the blue SPLASH ones had $D$ somewhat increased for
$z<1$, but remained basically unchanged for $z>1$ within the uncertainties.
The fractal dimensions of the quiescent galaxies had, nonetheless, the
most noticeable changes. For $z<1$ the red COSMOS2015 had bigger values
for $D$ in all distances measures, but suffered a considerable fractal
dimension reduction for the $z>1$. This same considerable decrease in
$D$ also happened for the red SPLASH galaxies in both ranges, $z<1$ and
$z>1$. Finally, the theoretical prediction for the reduction of the
fractal dimension in the range $z>1$ is also observed in all cases and
for all distance measures.

\section{Conclusions}\lb{conclusion}

This paper extended the study of \cite{teles2020} by applying the
same fractal analysis methodology to much larger galaxy samples 
in order to empirically test if large-scale galaxy distributions
can be described as fractal systems and if galaxy types, however
they are defined or observed, could possibly be dependent on the
single fractal dimension $D$. Tools originally developed for
Newtonian hierarchical cosmology were extended and applied to
relativistic cosmological models in order to describe possible
galaxy fractal structures by means of $D$ at deep redshift scales.
These tools were applied to the COSMOS2015 and SPLASH galaxy
survey datasets comprising almost one million objects spanning the
redshift interval of $0.1\le z\le6$.

In order to obtain volume-limited subsamples, absolute magnitudes 
were calculated using the measured redshifts in order to obtain the
respective luminosity distances $\dl$ by assuming the $\Lambda$CDM
relativistic cosmological model and the apparent magnitude limit of
24.7 in the K-band. Then graphs of absolute magnitudes in the K-band
versus redshifts were plotted using three values for the Hubble
constant, $H_0=(65,70,75)$ km/s/Mpc. Objects whose absolute magnitudes
were above the respective apparent magnitude limit were disregarded,
as well as those having $z>4$. This procedure provided two subsamples
with about 402k objects, the first containing 230705 COSMOS2015
galaxies, and the second containing 171548 SPLASH ones. Fractal
analysis was then performed in these two subsamples 

As relativistic cosmologies have several definitions of observed
distance \citep{ellis2007}, only three distinct ones were used here,
namely $\dl$, the redshift distance $\dz$ and the galaxy area distance
$\dg$, also known as transverse comoving distance. The use of several
cosmological distance measures comes from the fact that relativistic
effects become strong enough for redshift ranges larger than $z\gtrsim
0.1-0.2$ \citep{juracy2008}, so these distance definitions produce different
results for the same redshift value at those ranges. An algorithm for
sorting the data was performed so that graphs of number densities vs.\
relativistic distances were plotted. Straight lines were then fitted
to the data in two scales, $z<1$ and $1\leq z\leq4$, whose slopes
allowed direct determination of the respective single fractal dimensions.

The results indicated two consecutive redshift ranges behaving as
single fractal structures in both catalogs. Rounding them off and
their respective uncertainties we found that for $z<1$ the COSMOS2015
galaxies produced $D=1.4\pm0.2$, whereas the SPLASH galaxies yielded
$D=1.0\pm0.1$. For $1\leq z\leq4$ the respective calculations produced
$D=0.5\pm0.3$ and $D=0.8\pm0.4$. These results were found to be
unaffected by changes in the Hubble constant within the assumed
uncertainty. In addition, no transition to observational homogeneity
was found in the data.

Subsamples were created by selecting blue star forming galaxies and
red quiescent ones from both the COSMOS2015 and SPLASH surveys. These
subsamples were subsequently filtered by the same absolute magnitude
limits applied to the unselected samples, resulting in datasets that
were used to generate number densities and then to calculate fractal
dimensions. The results showed that up to two decimal digits the
fractal dimensions of blue galaxies remained basically unchanged,
whereas some red galaxies showed noticeable reduction at the same
precision, especially the red SPLASH galaxies at both ranges $z<1$
and $1\leq z<4$. This indicates that the fractal dimensions of both
surveys are dominated by blue galaxies.

These results suggest that besides being descriptors of galaxy
distributions, the fractal dimensions could also be useful tools as
tracers of galaxy types and evolution. For this purpose the galaxy
number densities could be used directly as depicted here to obtain
fractal dimensions or, complementarily, by means of their respective
power spectra \citep{lrs2022}, in order to study distributions
consisting of several galactic types, however they are defined or
observed, and in different environments.

Therefore, generating subsamples of blue and red galaxies showed not
only that the overall theoretical expectations for the fractal
dimension remain valid at $z>1$, but also that the difference between the
behaviour of blue versus red galaxies indicate that the fractal dimension
can reliably be used as a tool to characterise populations of different
types in galaxy formation. So, one could think of further subsample
separations in order to study certain types of evolving galaxies or other
objects based on their fractal dimension. Under this perspective the fractal
dimension would be seen as an intrinsic property of the distribution of
objects in the Universe that could be modulated depending on the types of
galaxies formed, which means that $D$ can be used as a tool to study
different galaxy populations.

It is also important to emphasize that the results
reached here are dependent on the way this paper defines homogenity
and its possible detection. As noted above the FLRW cosmology is
homogeneous by construction in a purely geometrical sense, and as
observations go further along the past light cone the spatial
homogeneity of the model cannot possibly be observed \textit{in a
cumulative manner} at larger scales because the model becomes
increasingly observationally inhomogeneous even using almost all
distance measures to derive observational density \citep{ribeiro2001b}.
At small redshifts this effect is not noticeable because the present
time hypersurface superposes on the past light cone, but at ranges
$z\gtrsim 0.1-0.2$ they start to differentiate, rendering this
distinction observable in principle \citep{juracy2008}.

One way of detecting this observational inhomogeneity
is by using the set of tools advanced by the pioneers of hierarchical
(fractal) cosmology \citep{vaucouleurs70,wertz70,wertz71,pietronero87}
once they are appropriately extended to the relativistic setting,
namely, using observational distances, radially observed cumulative
number counts and radial densities derived from the latter. The
cumulative number counts is connected to the single fractal dimension
$D$ which is then used either as tracer of this observational
inhomogeneity, or the evolution of galaxy types, or both as have been
done in this paper.

The tension mentioned above about a possible transition
to homogeneity in the large-scale galaxy distribution can then be
attributed to the application of different concepts, tools and methods
for dealing with this issue. For instance, \citet{scrimgeour2012} and
\citet{goncalves2021} analyzed their data using the mean number of
galaxies in spheres up to a certain comoving distance, the so-called
counts-in-sphere, a concept quite different from the radial cumulative
number counts $\Nobs$ used here to define our key quantity $\gobs^\ast$
as shown in Eqs.\ (\ref{gobs-ast}) and (\ref{Nobs}). Hence, different
definitions and methods lead to different results. This should come as
no surprise because as it happens with cosmological distances there is
no unique way to define homogeneity in cosmology. Each definition and
its related methodology produce their own tools which lead to different
conclusions about their particularly adopted concept of homogeneity.

Finally, the results presented here raise similar questions as discussed
in \cite{teles2020}, which are why there is such a significant decrease
in the fractal dimension for redshift values larger than unity. This
could be an observational effect caused by data bias, due to the simple 
fact that many galaxies located beyond $z=1$ are not detected, reducing
then the observed galaxy clustering and, therefore, the associated
fractal dimension. Other possibility is of some bias associated to
the small angular areas of these surveys such that they would not
yield representative measurements of the entire sky distribution.

It has been previously thought that different cosmological parameters
could affect these results, but this possibility no longer appears
plausible since we have used here different values for the Hubble
constant in the calculations, and that only altered $D$ very slightly,
even so within the obtained uncertainties. This suggests that fractal
dimension results are robust to changes in the cosmological model,
at least as far as FLRW, or FLRW like, cosmologies are concerned.

Apart from possible observational biases, one might also attribute
the decrease in the fractal dimension to real physical effects.
Under this viewpoint galaxy evolution and large-scale structure
dynamics of the Universe are at play in causing such a decrease. So,
it is possible that $D$ changes with the redshift due to galaxy
evolution such as selection effects or, perhaps, change in the
anisotropy distribution of the underlying cosmology. In this sense
the change in $D$ signifies that there might be indeed much less
galaxies at high $z$, meaning that the Universe was void dominated
at those epochs because galaxies would be much more sparsely
distributed and in reduced numbers. This possibility is not far
fetched, since for some time there has been a theoretical prediction
stating that the galaxy distribution fractal dimension must indeed
fall at larger scales because theory forecasts a sharp decrease in
the observed number density at $z>1$ \citetext{\citealp{ribeiro92b},
Fig.\ 1; \citealp{ribeiro95}, Figs.\ 1 and 3; \citealp{ribeiro2001b},
Fig.\ 2}. Hence, the observational shift in $D$ to smaller values in
terms of higher $z$ as reported here and in previous studies could be
interpreted as simply the empirical verification of this theoretical
prediction.

\section*{Acknowledgments}

We thank a referee for very helpful and useful suggestions which
improved this paper. We are also grateful to the editor for useful
comments. S.T.\ thanks the Universidade Federal do Rio de Janeiro
for a PIBIC scholarship. A.R.L.\ is grateful to Brazil's National
Council for Scientific and Technological Development (CNPq) for the
financial support.

\section*{Data Availability Statements}

The data underlying this article are available in COSMOS2015
and SPLASH databases, respectively at
\url{https://ftp.iap.fr/pub/from_users/hjmcc/COSMOS2015/}
and \url{https://www-users.cse.umn.edu/~mehta074/splash/}.
The datasets were derived from sources in the public domain:
\url{https://cosmos.astro.caltech.edu/page/photom} and
\url{http://splash.caltech.edu/public/catalogs.html}.

\section*{Competing Interests}

The authors have no competing interests to declare that are relevant
to the content of this article.



\bibliographystyle{h}
\bibliography{cosmo} 


\bsp	
\label{lastpage}
\end{document}